\documentclass[aps,prl,preprint,showpacs,amssymb,amsmath]{%
revtex4}
\usepackage{graphicx}
\begin{document}


\title{Distance Geometry: 
A Viewing Help for the Solid--Liquid Phase Transition
in Small Systems}


\author{P. Labastie}
%
\affiliation{Universit\'e de Toulouse, CNRS, UPS, Laboratoire Collisions
Agr\'egats R\'eactivit\'e, IRSAMC; F-31062 Toulouse, France}
\email[]{pierre.labastie@irsamc.ups-tlse.fr}

\date{\today}

\begin{abstract}
Distance geometry is the study of the arrangements of points in space using
only the mutual distances between them. The basic idea in this letter is to
use distance geometry for thermodynamics studies of small clusters in the
microcanonical ensemble. There are constraints on these distances, which
are shown to explain some
characteristic features of the caloric curve in very small clusters containing 3
or 4 atoms. We
anticipate that this approach could give a novel insight into the phase
transitions in larger clusters as well. During these studies, we have
established a very general and rather simple result for the Jacobian determinant
of the change of variables
from Cartesian coordinates to mutual distances, which is of wide
applicability in the $N$--body problem.
\end{abstract}


\maketitle



It is well known that most of the tractable problems of physics are such because
there exist a ``good'' choice of variables, which drastically simplifies
equations to be solved.
Such a good change of variable has not been found yet for the
$N$--body problem. Actually, for computational purposes, it has been
agreed that the Cartesian coordinates
are the less expensive coordinates because of the simplicity of the equations of
motions when written in terms of them, at least in classical mechanics. On
the other hand, those coordinates introduce unnecessary invariance (the
euclidean group), while physics describes mainly interactions between systems
characterized by their relative position. In this respect, mutual distances
should be enough to describe all physical processes. Another reason for
considering mutual distances as variables is the simplicity of the
potential energy when written in terms of them.
It might thus be worth the try to go as far as possible with these variables in
the studies of model systems described by pair potential. The work in this
letter is a start in this direction and primarily of exploratory nature. It aims
at bringing attention to some remarkable results (the simplicity of the
Jacobian of the transformation from Cartesian coordinates to mutual distances)
and to the ease of representation given by these variables.

Mutual distances have been considered by Lagrange in his study of the 3-body
problem. Several formulas for expressing various quantities of the system as a
function of distances alone are known. The basic reference for these is
Blumenthal's book \cite{blumenthal1953}. A more modern account is given in
chapter 4 of \cite{EDMBook}. Note that distance geometry is an invaluable tool
in the study of protein conformation, where the problem is to determine all the
distances between a set of atoms, knowing just a few of them, as given by NMR
or X-ray studies \cite{crippen2004,havel98}.

Several new directions can be explored and some results will be
presented in forthcoming publications.
In this letter, this change of
variable
is applied to the computation of the configurational density of states
(CDOS) of a small cluster of $N$ atoms interacting by a pair potential, in a
$D$ dimensional space for the special cases $N=D$ or $N=D+1$. In both cases,
the number of distances is equal to the number $s$ of independent variables
specifying a configuration: $s=DN-D(D+1)/2$. Note that for $N>D+1$, one has to
choose a subset of the distances as internal variables, which introduces
technical complexities not usefull for the purpose of this letter. The general
case will be presented in a forthcoming publication\cite{moibientot}. The CDOS
is the fundamental
quantity in classical thermodynamics. Its computation involves an
$s$-dimensional integral. We show below that, when
mutual distances are used as variables for an $N=D$-atom clusters, this
$s$-dimensional integral can be approximately reduced to $s$ convolution
products. This reduction is even exact at low energy. In any case, it is much
easier to compute than an $s$-dimensional integral.

The $N=D=3$ case is especially interesting because it is possible to draw a
representation of the constant energy surface in the distance space. The
accidents of the density of states as a function
of energy (or rather of its logarithm, the entropy $S(E)$) can be visualized and
given a simple interpretation on this drawing. It is well known that accidents
of $S(E)$ are related to so-called phase transitions in
clusters \cite{labastie1990,gross2000,chomaz2000,lyndenbell1999}. The
solid--liquid transition (actually a phenomenon akin to this transition) has
been observed both in simulations \cite{berry1987,labastie1990,wales1996} and in
experiments \cite{schmidt1997,schmidt1998,gobet2002,jarrold2003,jarrold2005b,
nous2008}. We anticipate that the simple view given in this letter could be used
in understanding the solid--liquid transition in larger clusters as well.

This letter is organized as follows. First, we recall the formulas for the
microcanonical thermodynamics of a set of atoms and the definition of the CDOS.
Then, we make some considerations on the change of variables, and
give the (very simple in our case) formula for the Jacobian determinant. We next
give some
properties of the space of distances. We then present drawings of the constant
energy surface in this space in the case of $N=3$ and explain the qualitative
features of the entropy curve. Finally, we show how the CDOS can be expressed
as a convolution product and present a comparison with more conventional
calculations.

In the microcanonical ensemble, the system size (volume $V$, number of
constituents $N$) and energy $E$ are fixed. The fundamental function is the
density of states $\Omega(E,V,N)$. It exists for any finite system with a
classical Hamiltonian $H(\mathsf{P},\mathsf{X})$. Here, the notation
$\mathsf{X}$ represents the set of Cartesian coordinates of the constituents
and $\mathsf{P}$ the corresponding momenta.
The Hamiltonian $H$ is the sum of the
kinetic energy $T(\mathsf{P})$ and the potential energy $E_p(\mathsf{X})$, so
that the density of states is actually the convolution of the kinetic density of
states and the configurational density of states $\Omega_c(E,V,N)$
\begin{equation}
\Omega_c(E,V,N)=\frac{\partial}{\partial E}\int_{E_p(\mathsf{X})\le
E}\mathrm{d}^{3N}\mathsf{X}.\label{CDOS}
\end{equation}
Since the kinetic density of state can be computed analytically, the
configurational density of state is the quantity we focus on. The
configurational entropy $S_c(E)$ is defined by Boltzmann formula
\begin{equation}
S_c(E,V,N)=k_{_B}\ln\Omega_c(E,V,N)\label{boltzmann},
\end{equation}
with $k_{_B}$ the Boltzmann constant.
In the
remaining of the paper, the configurational entropy will be called just entropy
for short.

It can be shown \cite{ruelle1988} that, under mild conditions on the potential
energy function $E_p(\mathsf{X})$, the entropy is extensive in the thermodynamic
limit, and that in this limit, the second derivative of the entropy is always
negative. This is not true in a finite system: the entropy is not extensive
because surface effects are important, and furthermore it can show portions of
positive curvature, which are the accidents of the CDOS we mentioned above.

We now describe shortly the change of variables from Cartesian coordinates to
mutual distances. The Cartesian coordinates of an atom $a, (a=1\dots N)$, are
noted $x_{ia},(i=1\dots D)$. Rather
than the distances, it is more convenient to define the
variables $y_{ab}$ by
\begin{equation}
y_{ab}=\sum_{i=1}^D\frac{(x_{ia}-x_{ib})^2}{2}.
\end{equation}
When going from Cartesian coordinates to internal
coordinates, one should define a ``moving frame'' (see for example
Ref.~\onlinecite{littlejohn1997}). Amazingly enough, the Jacobian determinant
for the
transformation
does not depend on this choice.  We suppose that the cluster is contained in a
(hyper)spherical volume of radius $R$, with the center of mass fixed at
the center of the sphere. This eliminates the need to integrate on the
translation degrees of freedom. The integration on the rotation coordinates
gives a constant irrelevant multiplicative factor, the volume of the group
$SO(D)$, noted $\mathcal{V}_{\mathrm{rot}}$. Eventually, it can be shown that
the configurational volume element can be written
\begin{equation}
\int_{\mathrm{rotation}}\prod_{i,a}\mathrm{d}x_{ia}=\mathcal{V}_{\mathrm{rot}}
\,\mathcal{J}\,\prod_{1\le a<b\le N}\mathrm{d}y_{ab},
\end{equation}
with $\mathcal{J}=1$ if $N=D$ and $\mathcal{J}=1/V$ if $N=D+1$, where $V$ is the
volume of the parallelotope built on the $D+1$ points. This volume can be
computed as the square root of a determinant involving the $y_{ab}$'s
only \cite{blumenthal1953}:
\begin{equation}
V^2=-\det\left|\begin{array}{cccccc}
0&1&1&1&\cdots&1\\
1&0&-y_{12}&-y_{13}&\cdots&-y_{1N}\\
1&-y_{12}&0&-y_{23}&\cdots&-y_{2N}\\
1&-y_{13}&-y_{23}&0&\cdots&-y_{3N}\\
\vdots&\vdots&\vdots&-\vdots&\ddots&\vdots\\
1&-y_{1N}&-y_{2N}&-y_{3N}&\cdots&0
\end{array}\right|.\label{Eq:volumecarre}
\end{equation}

Let us now think of the set of distances $y_{ab}$ as defining the Cartesian
coordinates of a point in an abstract space $\mathcal{D}$, which we call the
`Distance Space'. There are several conditions to be verified by this point.
They amount to ensure that all the determinants similar to that of
Eq.~(\ref{Eq:volumecarre}), for any subset of the $N$ atoms, are positive.
It can be shown (see e.g. Ref.~\onlinecite{EDMBook}) that these conditions
restricts the representative point in the distance space to be inside a convex
cone. In other words, $\mathcal{D}$ is a space with a frontier.
\begin{figure}
\includegraphics[width=5cm]{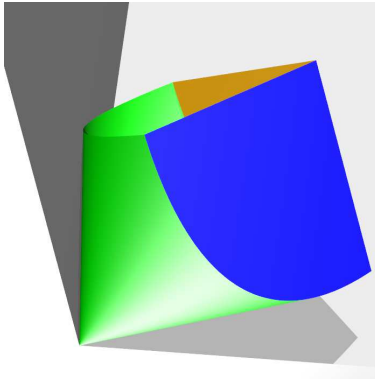}%
\caption{\label{figspace}(color on line) The distance space for $N=3$ atoms. The
axes for the coordinates $(y_{12},y_{13},y_{23})$ are shown by the
intersection of two planes. An acceptable point can only be inside the cone.
Furthermore, if the cluster is contained in a sphere, the accessible space is
further limited by 3 planes, which interset on the $(1,1,1)$ direction. On the
figure, one of the planes has been made transparent in order to show the
cone interior.}
\end{figure}

We have not yet included the condition that the atoms are confined inside a
sphere of radius $R$. Actually, the distance $r_a^2$ from atom $a$ to the center
of mass can be expressed in terms of the $y$ variables alone:
\begin{equation}
r_a^2=\frac{2}{N^2}\biggl[(N-1)\sum_{b\ne a}y_{ab}
-\sum_{\substack{b<c\\b,c\ne a}}y_{bc}\biggr].
\end{equation}
Imposing that $r_a^2\le R^2$ amounts to adding the further constraints that the
distance space is limited by a set of $N$ (hyper)planes. Those planes, together
with the cone defined above define a finite volume in the distance space. Figure
\ref{figspace} shows a view of the cone and the planes in the case $N=3$.

\begin{figure*}
\includegraphics[width=13cm]{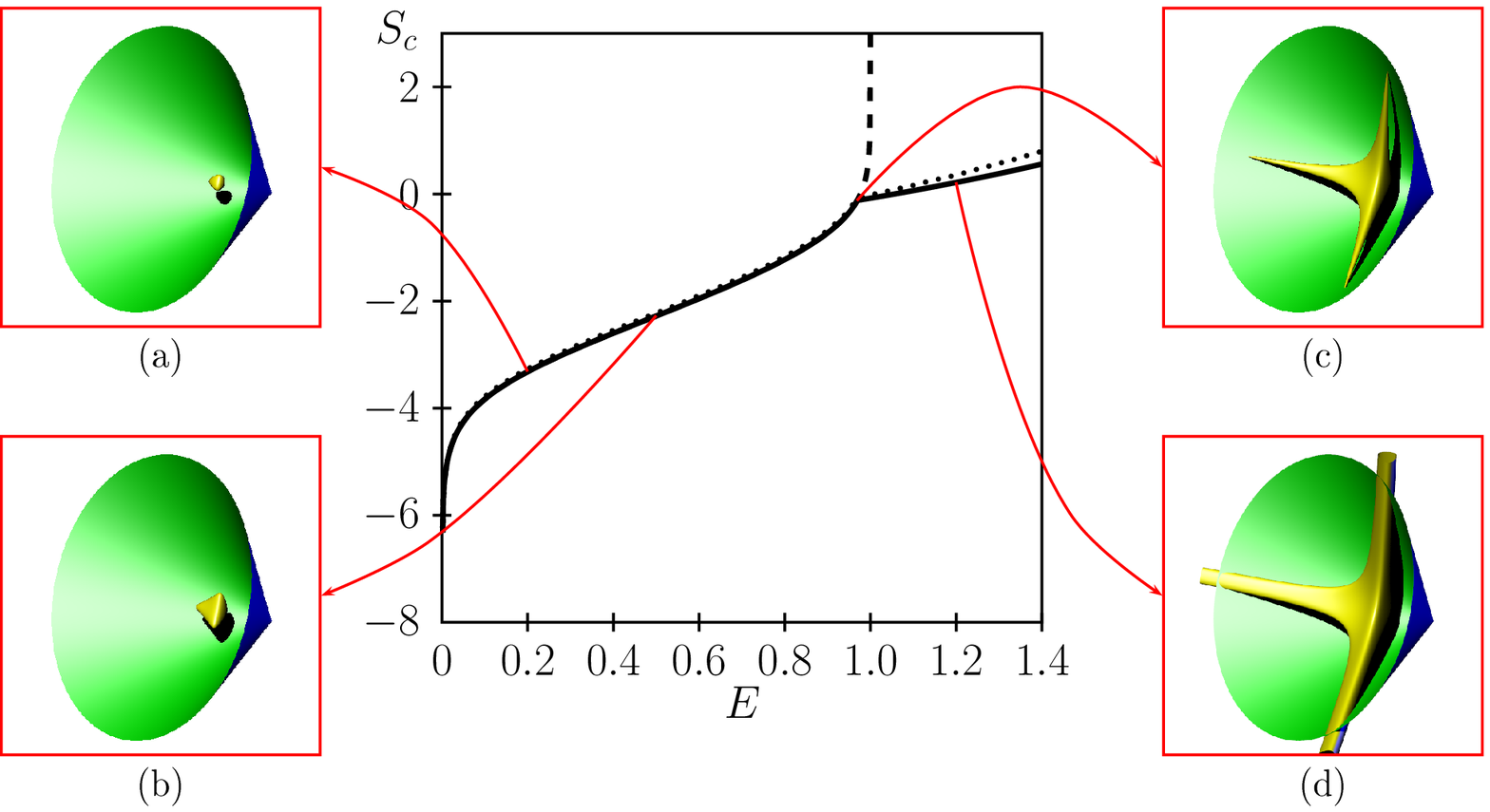}%
\caption{\label{figcone}Configurational Entropy $S$ as a function of energy $E$
for a 3 atom cluster interacting via Lennard-Jones potential. The insets
represent the CES in distance space (see text): (a) $E=0.2$,
(b) $E=0.5$, (c) $E=0.969$, (d) $E=1.2$. The dashed curve is the entropy
without constraint, the dotted curve is the result of a Monte-Carlo
calculation, the plain curve is the approximation by a convolution product (see
text).}
\end{figure*}

Now, What is the shape of the constant energy surface (CES) in distance space?
We suppose that
the potential is $E_p=\sum_{a\ne b}v(y_{ab})$, where $v(y)$ has the usual shape
of a pair potential: one minimum at $y=1$ with value -1, the limit when
$y\to\infty$ is 0 (that is the dissociation energy is 1) and the limit when
$y\to 0$ is $+\infty$. The CES in the case $N=3$ is
shown in the insets of Fig.~(\ref{figcone}). At low energy (inset
(a), $E_p=0.2$), the surface is close to a spheroid around the absolute minimum
of the
potential energy, which is $y_{12}=y_{13}=y_{23}=1/2$. Notice that this surface
is
far from the limiting cone. At higher energy (inset(b), $E_p=0.5$), the surface
becomes elongated in the direction parallel to the (distance) axes. More
precisely, it is easy to see that the intersection  of the surface  with the
axis $y_{12}=y_{13}=1/2$ for example, occurs for $y_{23}$ such that
$v(y_{23})=E_p$. This value of $y_{23}$ increases rapidly with increasing
$E_p$. There is of course a symmetry between the three axes. On the other hand,
at direction 45$^\circ$ from the axes, the maximum value of the $y$'s is such
that
$v(y)=E_p/2$, which is much closer to the point $(1/2,1/2,1/2)$. At an energy
close
to 1, the shape is that of 3 ``tentacles'' along the directions parallel to the
axis, and the surface reaches the enclosing cone for $E_p\simeq0.969$ (inset
(c)). This last value is specific of the Lennard-Jones potential; it is always
very close to one for any potential. For
$E>1$ (inset (d)), the CES is open. However, the volume
inside the cone is still finite. The configurational
entropy is shown on the same figure. At low energy, $S_c(E)\propto\ln(E)$,
which is the usual behavior. At higher energy, the entropy departs from the
logarithm, and would even diverge when $E\to1$ (the dashed curve). However, the
limitation imposed by the cone results in a discontinuity of the slope when the
CES meets the cone. It is important to understand that there is a balance
between the tendency of the pair potential to allow a distance to become large,
which results in a positive curvature of $S_c(E)$, and the limitation of pure
geometric nature due to the limiting cone, which bends in the negative
direction. Such ``bumps'' of the entropy curve are characteristic of a phase
transition in a finite system. We give here a new geometric view of what
happens, at least in this simple system.

\begin{figure}
\includegraphics[width=6cm]{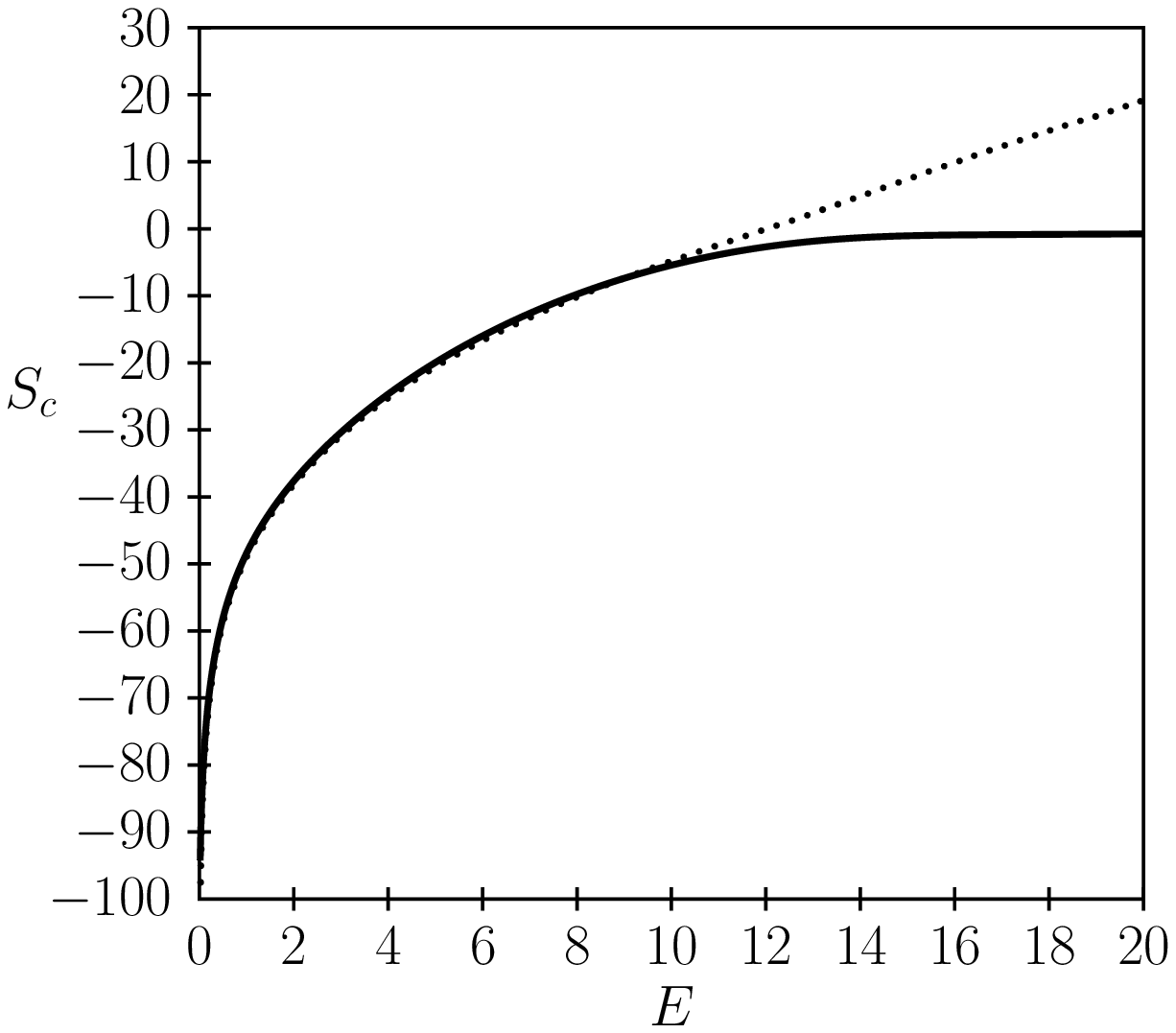}%
\caption{\label{fig8}Configurational Entropy $S$ as a function of energy $E$
for an 8 atom cluster interacting via Lennard-Jones potential. The dotted curve
is the result of a Monte-Carlo calculation, the plain curve is the approximation
by a convolution product.}
\end{figure}

Let us now compute the CDOS. The simplest case is $N=D$.
When $\epsilon<1$, the equation $v(y)=\epsilon$ has two solutions, which we
denote $y_{\mathrm{max}}(\epsilon)$ and $y_{\mathrm{min}}(\epsilon)$. When
$\epsilon\to1$, $y_{\mathrm{max}}\to\infty$ and disappears for $\epsilon>1$,
while $y_{\mathrm{min}}$ exists for any value of epsilon. To begin with, we
forget about any constraint on the distances, and we suppose that $N=D$. It is
easy to see that the CDOS can be written in this case
\begin{eqnarray}
  \Omega_c(E)&=&
\int_{\sum_{i=1}^s\epsilon_i=E}\biggl[\prod_{i=1}^s\omega(\epsilon_i)\,
\mathrm { d } \epsilon_i\biggr],\label{convolution}\\
\hbox{with}&&\omega(\epsilon)=\frac{\mathrm{d}}{\mathrm{d}\epsilon}
[y_{\mathrm{max}}(\epsilon)-y_{\mathrm{min}}(\epsilon)].\label{onedos}
\end{eqnarray}
Such an integral amounts to
$s-1$ convolutions. If the function $\omega$ is given on $n$ points, the
integral can be computed in $O(n^2s)$ time, to be compared to the $O(n^s)$
necessary for a general s-dimensional integral. Equation (\ref{convolution}) is
valid as long as the energy is low enough for the CES to not intersect the
limiting cone. Above that, an approximate density of states can be computed if
one supersedes Eq.~(\ref{onedos}) by
\begin{eqnarray}
\omega(\epsilon)&=&\frac{\mathrm{d}}{\mathrm{d}\epsilon}
[y_{\mathrm{max}}(\epsilon)-y_{\mathrm{min}}(\epsilon)]\quad\hbox{if
$y_\mathrm{max}(\epsilon)\le Y$,}\\
&=&\frac{\mathrm{d}}{\mathrm{d}\epsilon}
[\phantom{y_{\mathrm{max}}(\epsilon)}-y_{\mathrm{min}}(\epsilon)]
\quad\hbox{if
$y_\mathrm{max}(\epsilon)> Y$.}
\end{eqnarray}
$Y=(N-1)/N-2)$ is the value that zeroes the determinant (\ref{Eq:volumecarre})
with all the $y$'s equal to 1/2 except one, which is equal to $Y$. The result of
this approximation is shown on Fig.~\ref{figcone} for $N=3$ and Fig.~\ref{fig8}
for $N=8$. Notice that the good agreement with the exact CDOS extends far
beyond the expected range, which is $E<1$.

\begin{figure}
\includegraphics[width=5cm]{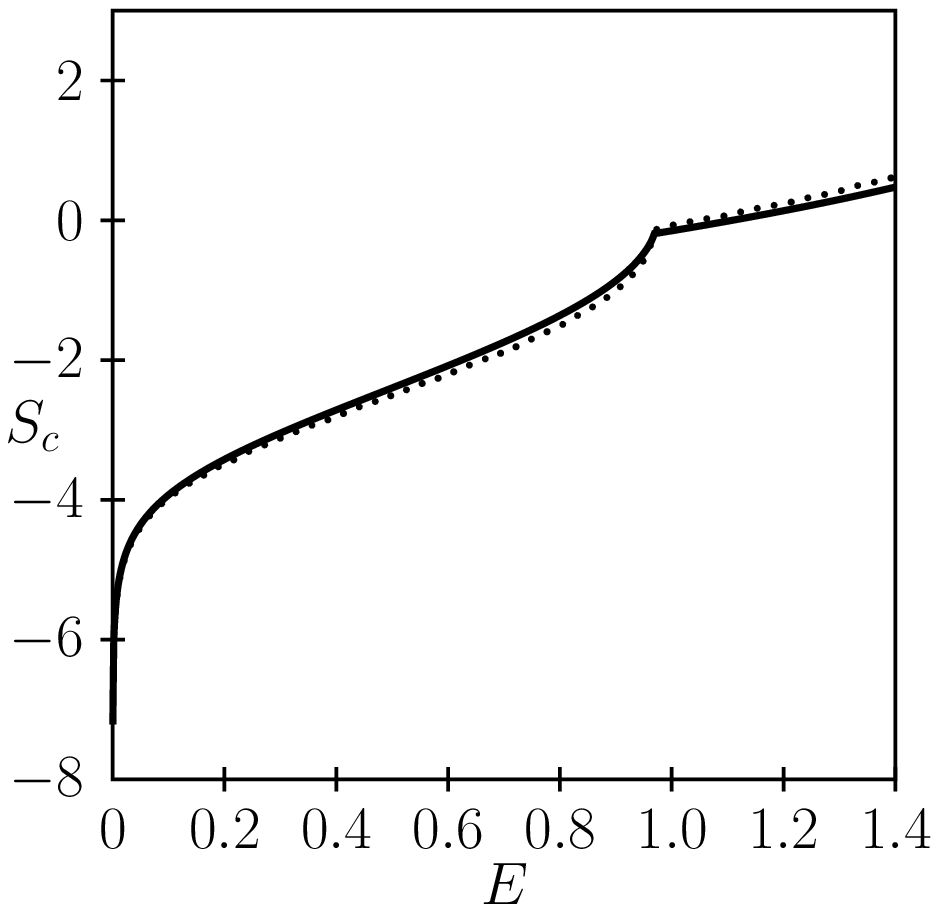}%
\caption{\label{figN3D2}Configurational Entropy $S$ as a function of energy $E$
for a 3 atom cluster interacting via Lennard-Jones potential in dimension 2.The
dotted curve
is the result of a Monte-Carlo calculation, the plain curve is the approximation
from Eq.~(\ref{approx}).}
\end{figure}

When $N=D+1$, the integral in Eq.~(\ref{convolution}) contains one more factor,
the Jacobian determinant, so that it is no more a convolution. On the other
hand, the CES develops essentially along directions parallel to the axes, hence
all the $y$'s are close to $1/2$. It can thus be included in the integral by
changing one of the $\omega$'s with
\begin{equation}
   \omega'(\epsilon)=\frac{1}{V[y_\mathrm{max}(\epsilon)]}
\frac{\mathrm{d}y_\mathrm{max}}{\mathrm{d}\epsilon}-
\frac{1}{V[y_\mathrm{min}(\epsilon)]}
\frac{\mathrm{d}y_\mathrm{min}}{\mathrm{d}\epsilon},\label{approx}
\end{equation}
where $V(y)$ is the volume of Eq.~(\ref{Eq:volumecarre}) with all the $y_{ab}$'s
equal to 1/2, save for one, which is $y$. Of course, when
$y_\mathrm{max}>Y$, only the term containing $y_\mathrm{min}$ is retained as
above. This approximation together with the exact Monte-Carlo result is shown
on Fig.~\ref{figN3D2} for 3 atoms in dimension 2. The agreement is still very
good, although the approximation is somewhat crude.

As a conclusion, 
mutual distances have been used as variables for the study of some
$N$-body problems. It has been shown that it is possible
to make efficient approximations for the calculation of classical densities of
states and that considering the space of distances is a good help for viewing
what happens at phase transitions in small systems.

\begin{acknowledgments}
I thank Ms. Ch.~Lauzeral for help on the initial stages of this project. Many
calculations have been done or checked with the help of the Maple software 
from Maplesoft, a division of Waterloo Maple Inc.
\end{acknowledgments}

\bibliography{letter030309}

\end{document}